\documentclass{article}
\usepackage{spconf,amsmath,graphicx}
\usepackage{hyperref}
\usepackage{bm}
\usepackage{multirow,multicol, makecell}
\usepackage{booktabs}
\usepackage{bbding}

\title{M$\bm{^2}$-CTTS: End-to-End Multi-scale Multi-modal Conversational \newline 
Text-to-Speech Synthesis
}
\name{%
\begin{tabular}{@{}c@{}}
Jinlong Xue$^1$,
Yayue Deng$^1$,
Fengping Wang$^1$,
Ya Li$^{1,*}$,
Yingming Gao$^1$,
Jianhua Tao$^2$,\\
Jianqing Sun$^3$,
Jiaen Liang$^3$ \thanks{* Ya Li is the corresponding author.}
\end{tabular}}

\address{
  $^1$Beijing University of Posts and Telecommunications, Beijing, China\\
  $^2$NLPR, Institute of Automation, Chinese Academy of Sciences, Beijing, China\\
  $^3$Unisound AI Technology Co., Ltd, Beijing, China}
  
\begin{document}
%
\maketitle
\begin{abstract}
    
 Conversational text-to-speech (TTS) aims to synthesize speech with proper prosody of reply based on the historical conversation. However, it is still a challenge to comprehensively model the conversation, and a majority of conversational TTS systems only focus on extracting global information and omit local prosody features, which contain important fine-grained information like keywords and emphasis. Moreover, it is insufficient to only consider the textual features, and acoustic features also contain various prosody information. Hence, we propose M$^2$-CTTS, an end-to-end multi-scale multi-modal conversational text-to-speech system, aiming to comprehensively utilize historical conversation and enhance prosodic expression. More specifically, we design a textual context module and an acoustic context module with both coarse-grained and fine-grained modeling. Experimental results demonstrate that our model mixed with fine-grained context information and additionally considering acoustic features achieves better prosody performance and naturalness in CMOS tests.

\end{abstract}
\begin{keywords}
speech synthesis, conversational TTS, prosody, multi-grained, multi-modal
\end{keywords}
\vspace{-0.2cm}
\section{Introduction}
\label{sec:intro}
\vspace{-0.1cm}

\begin{figure*}[t]
  \centering
  \includegraphics[width=\linewidth]{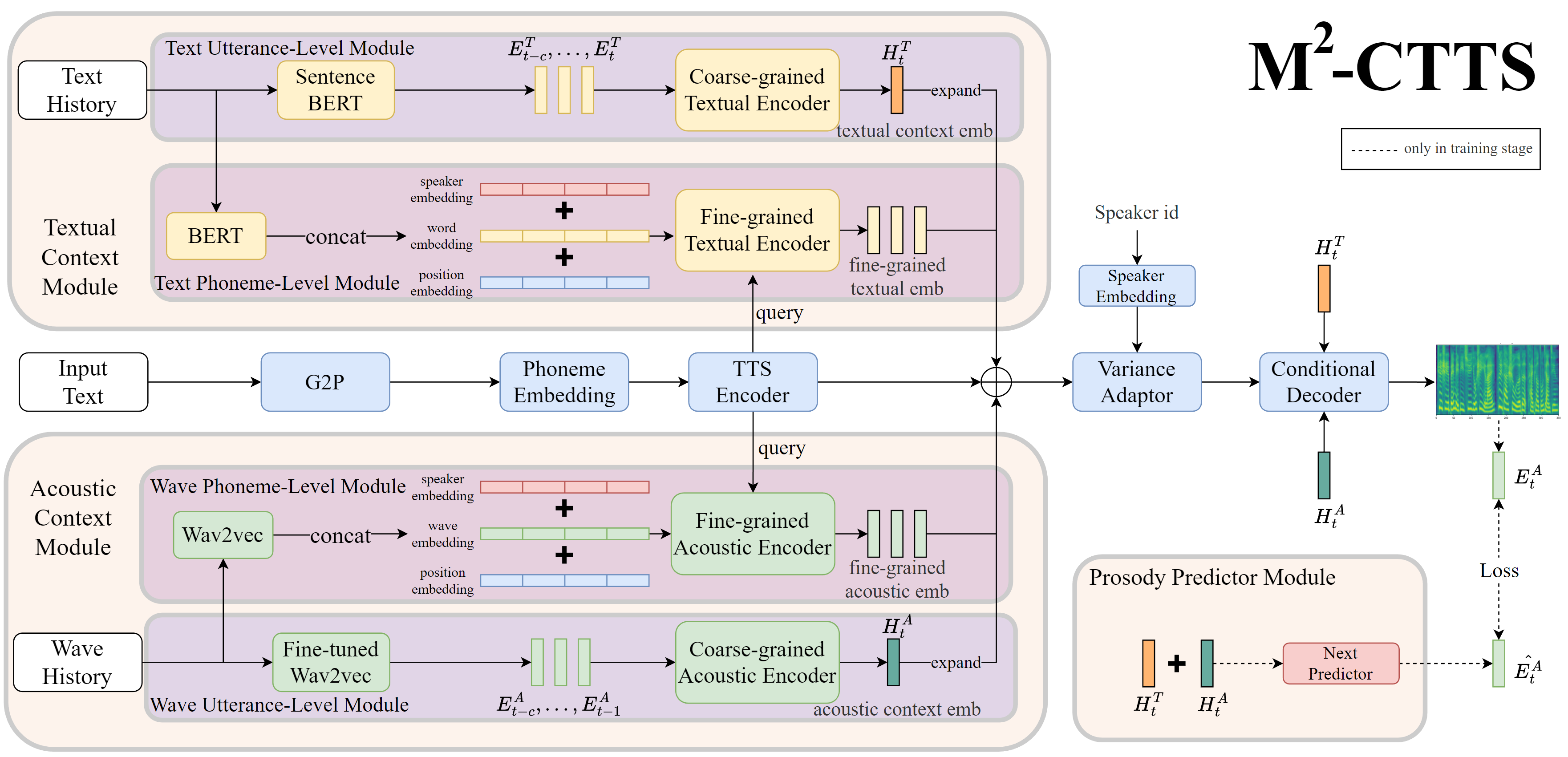}
  \caption{
  Overview of our proposed M$^2$-CTTS model. 
  }
  \label{fig:model_illustration}
\end{figure*}

    In the past few years, deep learning based text-to-speech (TTS) models have developed rapidly and they~\cite{skerry2018tacotron2, ren2020fastspeech2} can produce natural speech with narrator-like high quality that matches human levels. Moreover, by training on large-scale emotional speech corpus, many expressive speech synthesis systems have been proposed and they~\cite{wang2018styletokens,lee2019robust,min2021meta,wu2021cross,DBLP:journals/taslp/LeiYWX22} can produce speech with rich prosody and expressiveness. Conversational TTS, as a more complex technique in the field of Human-Computer Interaction (HCI), aims to synthesize speech with suitable prosody according to history conversation. Those expressive speech synthesis methods cannot perform well if they are directly applied to conversational TTS~\cite{DBLP:conf/icassp/LiMLWMWS22}, because conversational TTS relies heavily on historical conversation and it is crucial to model such dialogue information.   
     
    Some conversation modeling methods have been proposed recently. Guo et al.~\cite{DBLP:conf/slt/GuoZSHX21} proposed a context encoder to reinforce the utterance-level information of history context from textual aspect, and an extra auxiliary encoder is used to extract useful statistical text features like semantic and syntactic features. Cong et al.~\cite{DBLP:conf/interspeech/CongYHLX021} introduced an extra acoustic context encoder to acquire the acoustic embedding from previous speech. Besides, Li et al.~\cite{DBLP:conf/icassp/LiMLWMWS22} proposed a graph-based multi-modal context modeling method to model inter-speaker and intra-speaker influence in conversations.
    
    However, these researches in conversational TTS only focus on predicting prosody embedding from the global level within single modality and neglect to consider local features from different aspects, which are insufficient to reflect actual human behaviors~\cite{xie2021encoding} like keywords emphasis or changes of prosody. Hence, we adopt a hierarchical modeling method, including utterance-level and phoneme-level modules from both textual and acoustic aspects. 
    Moreover, some methods~\cite{DBLP:conf/slt/GuoZSHX21,DBLP:conf/interspeech/CongYHLX021} simply use concatenation operation and GRU for modeling, which has a limited ability to describe relationship between current and past utterances. Hence, we enhance the connection by multi-head cross-modal attention mechanism in our work. Besides, we further apply next predictor module for better prosody similarity and use Style-Adaptive Layer Normalization (SALN)~\cite{min2021meta} to infuse prosody information. We conduct experiments on conversational speech corpus DailyTalk~\cite{lee2022dailytalk} and the results show that our method outperforms other methods in terms of CMOS test. Especially, fine-grained expressiveness has been greatly strengthened.
    
    Our major contributions are listed in the following. (1) we not only extract global features but also make use of local information to enhance expressiveness, and (2) we develop a method that uses both textual and acoustic conversation information, and (3) we better utilize prosody embeddings by applying SALN and prosody predictor module. 
    
    The rest of the paper is organized as follows. Section~\ref{sec:c3m} introduces our proposed conversational TTS system in detail. Section~\ref{sec:experiments} shows our experiment results and discussions. The paper is concluded in Section~\ref{sec:conclusions}.


\vspace{-0.1cm}
\section{M$^2$-CTTS Model}
\label{sec:c3m}
\vspace{-0.1cm}


    
    The entire model architecture of M$^2$-CTTS is depicted in Fig.~\ref{fig:model_illustration}. M$^2$-CTTS is based on FastSpeech2 \cite{ren2020fastspeech2} which consists of an encoder, a decoder and a variance adaptor. In order to capture both textual and acoustic information from the history conversation, we design a Textual Context Module (TCM) and an Acoustic Context Module (ACM). Furthermore, in order to consider hierarchical history context at the same time, we design a Text Utterance-Level Module (TUM) and a Text Phoneme-Level Module (TPM) in TCM and design a Wave Utterance-Level Module (WUM) and a Wave Phoneme-Level Module (WPM) in ACM. Moreover, we use Conditional Decoder and Prosody Predictor Module (PPM) to enhance the expressiveness of the generated speech. Specifically, we adopt Style-Adaptive Layer Normalization (SALN)~\cite{min2021meta} in decoder as Conditional Decoder to increase the expressiveness and better utilize the history context. 

\vspace{-0.2cm}
\subsection{Multi-scale Dialogue Modeling} 
\label{sec:Multi-scale}

    Although global information like emotions or intentions allows the model to understand the conversation, detailed information like keywords or emphasis that include strong or implicit meanings should also be highlighted. Besides, inconsistent or mixed style within utterances may lead to unnatural prosody. We also empirically found that fixed-length conversation embedding worked poorly and led to average prosody. Hence, we additionally combine coarse-grained with fine-grained conversation modeling and adopt multi-head attention to acquire comprehensive history information. We denote the conversation history at $t$ time stamp as:
    \begin{equation}
    \vspace{-0.1cm}
    Conversation = \{ A_{t-c}, B_{t-c+1}, …, A_{t-1}, B_{t} \}
    \label{conversation_process}
    \end{equation}
    where $c$ is a memory capacity parameter that determines how many turns are taken into consideration between two speakers A and B.

    \textbf{Coarse-grained Context Modeling.} In terms of coarse-grained context modeling, we utilize Text Utterance-Level Module (TUM) and Wave Utterance-Level Module (WUM) to acquire global context information.
    
    For TUM, we follow and modify the module proposed in~\cite{DBLP:conf/slt/GuoZSHX21} to model the textual utterance-level information, which consists of a Sentence BERT~\cite{reimers2019sentencebert} based textual utterance-level feature extractor and a Coarse-grained Textual Encoder. Because Sentence BERT can capture semantic information, we adopt it to extract history sentences embeddings $E^T_{t-c:t-1}$ from $Conversation$ with last $c$ turns dialogue and current utterance embedding ${E^T_t}$.  
    For Coarse-grained Textual Encoder, a GRU layer is used to encode the history embeddings $E^T_{t-c:t-1}$ first for considering cumulative past conversational information. Then we concatenate the final hidden states of GRU with the current utterance embedding ${E^T_t}$ and feed them into a linear layer with additional attention mechanism to calculate the weighted global textual context embedding $H^T_t$ as shown in Fig.~\ref{fig:model_illustration}. 
    
    For WUM, we adopt a similar architecture to extract the global acoustic context embedding $H^A_t$, except for the feature extractor and we do not use the speech of the current utterance $B_t$. Instead of the conventional methods that use spectral-based features and a reference encoder or Global Style Token (GST)~\cite{wang2018styletokens}, we adopt a Wav2vec 2.0~\cite{baevski2020wav2vec} fine-tuned in the downstream Speech Emotion Recognition (SER) task to extract the acoustic utterance-level prosody information $E^A_i$ directly from raw audio. This method brings twofold benefits. The pretrained model can capture more comprehensive acoustic features in contrast to spectral-based features which lack phase information and contain less prosodic data than raw audio. In addition, using pretrained model can mitigate the training difficulty of complex models and have robust performance.

    \textbf{Fine-grained Context Modeling.} Although coarse-grained context modeling module can capture information from the whole utterances or connections between utterances, it is unnatural to give every phoneme the same prosodic embedding which results in plain prosody of generated speech. Considering a real-world conversation, people will react to specific words or phrases that others say in communication. Therefore, in order to simulate these situations, we develop a Text Phoneme-level Module (TPM)  and a Wave Phoneme-Level Module (WPM) to model fine-grained history information. 
    
    In TPM and WPM, hidden feature sequence $P_i$ from i-th sentence or utterance is acquired from BERT~\cite{devlin2018bert} and Wav2vec 2.0~\cite{baevski2020wav2vec} respectively. Multi-head cross-modal attention mechanism is then utilized to identify essential phrases or frames which have an important impact on context understanding and expressiveness. A more detailed introduction is below. 
    

\vspace{-0.2cm}    
\subsection{Multi-modal Context Modeling}
\label{sec:Multi-model}
    In addition to semantic and syntactic features, paralinguistic information like acoustic prosodic cues also play an important role in context understanding, since people speak in different tones even in the same contents or situations. Thus, modeling both modalities is significant.  
    
    \textbf{Textual Context Modeling.} As shown in Fig.~\ref{fig:model_illustration}, Textual Context Module contains TPM and TUM. TPM consists of a feature extractor and a Fine-grained Textual Encoder.
    We adopt BERT~\cite{devlin2018bert} to acquire the hidden feature sequences $P_i$ from i-th sentence and aggregate all last $c$ dialogues into a long sequence.
    After that, we add the speaker ID to reserve speaker identity and we also use sinusoidal positions to represent the conversation orders. In terms of Fine-grained Textual Encoder, 1D convolutional layer is first adopted for contextualization. The output of TTS encoder is then used as the query for a multi-head cross-attention module with historical fine-grained representation sequences as both key and value. The resulting weighted context representation will be added back to the output of TTS encoder with the residual design. The structure of TUM is depicted in Sec.~\ref{sec:Multi-scale}. Thus, the important keywords or phrases from the textual aspect are considered.

    \textbf{Acoustic Context Modeling.} As shown in Fig.~\ref{fig:model_illustration}, Acoustic Context Module includes WPM and WUM. WPM consists of a feature extractor and a Fine-grained Acoustic Encoder. Different from using mel spectrogram as acoustic input, we adopt Wave2vec 2.0~\cite{baevski2020wav2vec} to extract the hidden feature sequences from raw audios. The other processes are similar to TPM described above. Therefore, the local expression or emphasis from the acoustic perspective is noticed.
  
\vspace{-0.2cm}
\subsection{Constraint of Prosody}
    Similar to~\cite{DBLP:conf/interspeech/CongYHLX021}, we additionally utilize a Prosody Predictor Module (PPM) to predict prosody embedding of current utterance $E_t^A$ from multi-modal global context embeddings $H_t^T$ and $H_t^A$. As shown in Eq.~\ref{loss}, an additional MSE loss is adopted to constrain coarse-grained context modules to predict the prosody expression of current utterance based on the history.
    \begin{equation}
        \hat{E}_t^A = PPM(H_t^T, H_t^A), \  Loss = || \hat{E}_t^A, E_t^A ||
    \label{loss}
    \end{equation}
    where $\hat{E}_t^A$ stands for predicted prosody embedding. $E_t^A$ is the ground-true prosody embedding extracted by fine-tuned Wav2vec model.
    It should be noted that the prosody predictor module and loss computation are only used during training. 
    
\vspace{-0.2cm}
\section{EXPERIMENTS}
\label{sec:experiments}
\vspace{-0.1cm}

\subsection{Experimental Setup}

    We use public English corpus DailyTalk~\cite{li2017dailydialog} to conduct our experiments. The DailyTalk dataset is derived from DailyDialog dataset~\cite{li2017dailydialog} and it contains 2,541 dialogues performed by one male and one female speaker. Every dialogue in DailyTalk has more than five turns. For convenience, they are split into 23,773 audio clips by dialogue turns and they contain 20 hours in total. We leave 128 dialogues for validation set and others for training set. In our experiments, all utterances are down-sampled to 22050Hz and are used to extract 80 dimensional mel spectrogram.
    
    We use FastSpeech2 as our TTS backbone and we adopt unsupervised duration modeling~\cite{badlani2022one} rather than using supervised duration model with external aligner, such as Montreal Forced Aligner~\cite{mcauliffe2017montreal}, because external aligners have risk of out-of-distribution problem and soft alignment has more flexibility for expressive performance. We use the pretrained HiFi-GAN~\cite{kong2020hifi} model as our vocoder to convert the 80 dimensional mel spectro-grams to 22050Hz audio files. Our conversational TTS models are all
    trained for 400K steps with a batch size of 16 on GeForceRTX 3090. The training hyperparameters such as optimizers are based on the original configuration.
    
    For textual feature extractors, we use pretrained Sentence BERT
    \footnote{https://huggingface.co/sentence-transformers/distiluse-base-multilingual-cased-v1}
    in TUM, and use the original BERT
    in TPM. For acoustic feature extractors, we adopt pretrained Wav2vec 2.0
    \footnote{https://huggingface.co/speechbrain/emotion-recognition-wav2vec2-IEMOCAP}
    fine-tuned on IEMOCAP~\cite{busso2008iemocap} emotional training data in WUM, and use original Wav2vec 2.0
    in WPM.
    
    To study the performance of our proposed method, we use the comparison methods M0 to M7 for evaluation. Detailed combinations are in Table~\ref{tab:mos}. M1 is the FastSpeech2 baseline model and M2 to M7 all adopt Conditional Decoder and prosody Predictor Module to enhance the expressiveness.

\begin{table}[t]
  \caption{The comparison methods and the MOS results.}
  \label{tab:mos}
  \centering
  \scalebox{.95}{
      \begin{tabular}{c|c c c c|c}
        \toprule
        Methods & TUM & WUM & TPM & WPM & MOS Score \\
        \hline
        M0 & \multicolumn{4}{c|}{reconstructed from ground-truth} & $4.38\pm 0.06$ \\
        \hline
        M1 &  &  &  &  &  $3.64\pm 0.07$ \\
        M2 & \Checkmark &  &  &  & $3.71\pm 0.06$ \\
        M3 &  & \Checkmark &  &  & $3.65\pm 0.07$ \\
        M4 & \Checkmark &  & \Checkmark &  & $3.64\pm 0.07$ \\
        M5 & & \Checkmark &  & \Checkmark  & $3.70\pm 0.07$ \\
        M6 & \Checkmark & \Checkmark &  &  & $3.73\pm 0.06$ \\
        M7 & \Checkmark & \Checkmark & \Checkmark & \Checkmark & $3.74\pm 0.07$ \\
        \bottomrule
      \end{tabular}
      }
\vspace{-0.2cm}
\end{table}

\vspace{-0.1cm}
\subsection{Evaluation}

     We conduct comparison mean opinion score (CMOS) test to evaluate the performance of each designed pair (A, B) in test set. We randomly select 20 synthesized utterances at i-th turns $U_i$ and four ground-truth history contexts with both text and speech $U_{i-4:i-1}$ for reference to compare the naturalness and expressiveness. Especially, participants need to judge whether the prosody is more expressive and more suitable with the history dialogue, and rate on a scale from -3 (Completely A model better) to 3 (Completely B model better) with 1-point discrete increments according to the criteria. The absolute value shows the degree.
     
     The CMOS results are shown in Table~\ref{tab:cmos}. We find that using TUM or WUM to model dialogue history in utterance-level from textual or acoustic aspects respectively can both improve the expressiveness a lot. We also find that combining fine-grained features in acoustic aspect has significant benefits in synthesizing speech with rich prosody. However, we do not find similar conclusion in textual aspect. 
     We suppose that local acoustic features contain more implicit information, since people utter the same word or phrase with a variety of prosody and emotion in different situations. 
     Comparison ``M4 vs. M5" considering multi-scale information but from different aspects also confirms that acoustic features is more effective. Furthermore, we investigate the performance of adopting both textual and acoustic modalities. CMOS score of ``M2 vs. M6" indicates that combining two aspects in utterance level can just slightly enhance the performance. We then visualize the attention and find that both TUM and WUM focus on the same utterances in most cases, especially the latest turn. As a consequence, the extracted information has high degree of similarity. The test cases ``M2 vs. M7" and ``M6 vs. M7" show that our method is superior to other methods~\cite{DBLP:conf/slt/GuoZSHX21,DBLP:conf/interspeech/CongYHLX021} respectively proposed in earlier research. M2-CTTS has better understanding of the history conversation and greatly enhances the fine-grained prosodic expression. 
     
     We also conduct a five-scale mean opinion score (MOS) test to evaluate the quality and naturalness of the speech. The greater the score, the better the quality. Results in Table~\ref{tab:mos} show that modeling the prosody information from different aspects can relieve the problem of generating average prosody. We also notice that despite our method containing many modules, it does not necessarily impair the quality and naturalness thanks to the prior knowledge from pretrained models. Examples of synthesized speech can be found on the project page~\footnote{Audio samples: https://happylittlecat2333.github.io/icassp2023}.

\begin{table}[t]
  \caption{The results of the CMOS tests. CMOS describes the preference degree between A and B (denote as A vs. B).
  The Preference is calculated according to the CMOS score.}
  \label{tab:cmos}
  \centering
  \scalebox{1.}{
      \begin{tabular}{c|c c c c}
        \hline \hline
                  &        & \multicolumn{3}{c}{Preference(\%)} \\
        \cline{3-5}
                  & CMOS   & Left    & Neutral    & Right   \\
        \hline
        M1 vs. M2 & 0.25   & 18.9    & 46.1       & 35.0    \\
        M1 vs. M3 & 0.19   & 14.4    & 53.9       & 31.7    \\ 
        M2 vs. M4 & 0.03   & 25.6    & 45.5       & 28.9    \\ 
        M3 vs. M5 & 0.21   & 17.2    & 46.1       & 36.7    \\ 
        M4 vs. M5 & 0.19   & 18.9    & 45.5       & 35.6    \\ 
        M2 vs. M6 & 0.05   & 25.6    & 43.8       & 30.6    \\ 
        M2 vs. M7 & 0.34   & 17.8    & 38.3       & 43.9    \\ 
        M6 vs. M7 & 0.32   & 16.7    & 41.1       & 42.2    \\ 

        \hline \hline
      \end{tabular}
  }
\vspace{-0.3cm}  
\end{table}

\vspace{-0.3cm}
\section{CONCLUSION}
\label{sec:conclusions}
\vspace{-0.1cm}

    In this paper, we propose an end-to-end multi-scale multi-modal conversational speech synthesis system (M$^2$-CTTS) that models multi-grained context information extracted from both acoustic and textual features. Experimental results show that considering both acoustic and textual modalities can enhance the prosody and naturalness of synthesized speech. Besides, combining coarse-grained features with fine-grained features will further improve expressiveness. especially the fine-grained prosody extracted from acoustic features. For future work, we will investigate the fine-grained modeling of conversational TTS with controllability.

\vspace{-0.2cm}
\section{Acknowledgements}
\label{sec:ack}
\vspace{-0.1cm}

This work is supported by the National Natural Science Foundation of China (NSFC) (No. 62271083) and Open Project Program of the National Laboratory of Pattern Recognition (NLPR) (No. 202200042).


\bibliographystyle{IEEEbib}
\bibliography{strings,refs}

\end{document}